# SIMULACION DE LA DINAMICA EN LA LAMINA DE CORRIENTE DE LA MAGNETOSFERA

Ojeda A., B. Lazo, A. Calzadilla, S. Savio, K Alazo

Departamento de Geofísica Espacial, Instituto de Geofísica y Astronomía (IGA), arian@iga.cu

## Resumen

La dinámica de la lámina de corriente de la magnetosfera fue simulada haciendo transformaciones a un arreglo rectangular N×M de celdas (autómata celular) originalmente propuesto por *Koselov and Koselova (2002)*. La parte de la magnetosfera del sistema modelado fue organizada como un arreglo rectangular de celdas con una energía acumulada y concurrirá una redistribución local de energía cuando se exceda un valor umbral en una de las celdas. Se asume que el valor umbral en cada celda depende de un parámetro de control externo que influye a lo largo de la frontera de la matriz (40x80). La dinámica del modelo es controlada por la componente z del Campo Magnético Interplanetario (Bz), además se discuten las analogías entre los procesos transciendes y el disparo de subtormentas aurorales. Los Bz corresponden a ventanas temporales del viento solar para un grupo de nubes magnéticas y plasmoides. El modelo simula patrones organizados en la distribución de energía. La función de distribución de probabilidad (o PDF) del tamaño de las avalanchas tienen forma de ley de potencia, mostrando evidencias de un sistema con Criticalidad Auto-Organizada (SOC). La ocurrencia de grandes avalanchas en el modelo está relacionada con el disparo de alguna subtormenta magnética intensa. Como la llegada a la magnetosfera terrestre de un viento solar con un Bz negativo es la condición fundamental para la ocurrencia de subtormentas magnéticas, mientras mayor sea el valor modular del Bz negativo, mayor es el valor del índice AE y se activa una fuerte subtormenta magnética.

## Abstract

### SIMULATION OF THE DYNAMICS IN THE MAGNETOSHEET CURRENT SHEET

The dynamics in the magnetosphere current sheet was simulated following transformations to the rectangular N×M array of cells (cellular automaton) originally proposed by *Koselov and Koselova (2002)*. The magnetosphere part of the modeling system was organized as a rectangular arrangement of cells with a stored energy, a local redistribution of the energy will exist when a value threshold is exceeded in one of the cells. We assume that the threshold value in each cell depends on external control parameter which influences the long boundaries of the rectangular array (40x80). The model dynamics controlled by the z-component of the interplanetary magnetic field (Bz) as well as analogies between the model transient processes and the observed substorm auroral activations are discussed. The Bz correspond to temporary windows of the solar wind for a group of magnetic clouds and plasmoids. The model simulates organized patterns in the energy distribution. The function of distribution of probability (or PDF) of the size of the avalanches have a power-law form, showing evidences of a system with Self-Organized Criticality (SOC). The occurrence of big avalanches in the model is related with the activation of some intense magnetic substorm. The arrival to the terrestrial magnetopause of a solar wind with Bz negative is a fundamental condition of occurrence of magnetic substorms. While more negative it is the Bz IMF component bigger it is the value of the AE index and trigger a strong magnetic substorm.

## Introducción

La actividad solar modula el clima terrestre, pues las reacciones de fusión que ocurren en el interior del Sol hacen que se genere energía, siendo una de sus formas la radiación electromagnética que barre casi todo el espectro de frecuencias y que llega a la Tierra en aproximadamente 8 minutos. De manera simultánea el Sol constantemente está expulsando plasma al medio interplanetario y su capa más externa, la corona solar, se expande más allá de los límites del sistema solar, este flujo constante de plasma es lo que se a dado en llamar, viento solar. El movimiento de este flujo de plasma tiene un carácter caótico e intermitente, en dependencia de la actividad solar y en general los valores promedios de su velocidad son de 450 Km/s, su densidad de 5 protones/cm$^3$ y su campo magnético de 5 nT.

El plasma solar tiene una elevada conductividad eléctrica y debido a ello arrastra consigo en su movimiento al campo magnético del Sol, creándose debido a la rotación solar, una distribución del Campo Magnético Interplanetario en espiral. Este Campo Magnético Interplanetario tiene una estructura sectorial, con regiones donde las líneas del campo parten del Sol y otras donde las líneas entran al Sol. De forma cuasi-periódica el Sol cambia la polaridad de su campo magnético, así como el nivel de su actividad, siendo medible esta última a partir de las distintas manifestaciones que tienen lugar en éste: aparición de manchas y/o regiones activas, variaciones del área de las regiones llamadas huecos coronales, etc.

En general estos ciclos solares van desde 9 años hasta casi 13, dando como promedio relativo una duración de aproximadamente 11 años. Durante estos períodos la variabilidad del viento solar y el Campo Magnético Interplanetario modulan en buena medida el clima terrestre, pudiéndose generar de la interacción entre ambos, fenómenos llamativos como las auroras polares y las tormentas geomagnéticas.

La radiación ionizante proveniente del Sol influye directamente en la ionosfera terrestre, ya que su estructura en capas es una consecuencia de la interacción de los constituyentes atmosféricos con la radiación ultravioleta, rayos gamma, radiación infrarroja y rayos x. El estudio de los fenómenos que se producen en el Sol y se propagan en el viento solar afectando la magnetosfera e ionosfera terrestres constituyen un reto científico permanente por la incidencia que tienen sobre la tecnología y el equipamiento para las comunicaciones en general y el posicionamiento global por satélites, ya que cada día la humanidad es mas dependiente de éstos.

Un fenómeno que ha sido muy estudiado y aun hoy es una región fértil para la investigación son las eyecciones de masa coronal, por la necesidad de realizar su caracterización para evaluar el grado de afectación que generaran al interactuar con la magnetosfera e ionosfera terrestre.

Algunas eyecciones de masa coronal se clasifican por sus características como nubes magnéticas, si cumplen una serie de propiedades definidas por Burlaga [*Burlaga et al., 1981, 1982; Schwenn and Marsch, 1991*]. Estos eventos cuando se propagan hacia la Tierra tienden a ser muy geoefectivos, pues traen un campo magnético que rota suavemente con un período de duración del orden de 24 horas y logra reconectarse con el campo magnético de la Tierra incorporando gran cantidad de partículas energéticas a la magnetosfera terrestre. Esto hace que se incremente la intensidad de las corrientes eléctricas en el interior de la magnetosfera y por tanto se induzcan campos magnéticos a nivel de superficie, produciéndose lo que se conoce como tormenta geomagnética.

Durante el desarrollo de una tormenta geomagnética crece el óvalo auroral, expandiéndose hacia las bajas latitudes. Estos fenómenos pueden afectar la actividad del hombre y causar pérdidas millonarias como el caso de la tormenta del 13 de marzo de 1989 que provocó la falta de fluido eléctrico en la ciudad de Québec, Canadá y al noreste de Estados Unidos, además se ha corroborado que pueden dañar satélites artificiales de comunicación, afectar las comunicaciones vía ionosfera, aumentar los niveles de corrosión en los oleoductos, entre otros.

Los Autómatas Celulares son estructuras ideales para construir modelos digitales aproximados de algunos sistemas complejos de naturaleza continua, sin pasar por modelos analógicos. Es posible, por ejemplo, lograr sencillos modelos digitales que representen con suma fidelidad algunas leyes de la Física. Un Autómata Celular es una herramienta computacional de la Inteligencia Artificial, basada en modelos biológicos o físicos, básicamente compuesto por una estructura estática de datos y un conjunto finito de reglas que son aplicadas a cada nodo o elemento de la estructura.

El interés que ha despertado esta técnica radica en la sencillez y en la simplicidad que caracteriza la construcción de los modelos; así como en la particularidad de los patrones de comportamiento presentados por el Autómata en tiempo de ejecución [*González Vargas, 1999*].

Según *Toffoli y Margolus (1987)*, se define un Autómata Celular sólo si:

- En cualquier instante de tiempo cada celda tiene una propiedad bien definida, o sea, todas tienen el mismo conjunto $\Sigma$ de estados posibles. Por ejemplo: En un modelo de incendio forestal, cada celda puede tener tres estados: un árbol verde, un árbol en llamas o un sitio vacío.
- Cada celda pasa de un Estado a otro por reglas bien definidas (a medida que pasa el tiempo) haciendo al sistema inteligente. Luego, las celdas tienen la misma Regla de Evolución.
- Cada celda tiene la misma cantidad de vecinos en posiciones simétricas o no, de los cuales conocemos su Estado, o sea, todas tienen la misma forma de Vecindad.

Un Autómata Celular puede ser construido definiendo alguna especificación para cada uno de sus componentes, lo que permitirá cierta flexibilidad en el momento de construir el autómata.

- El autómata puede ser de 1, 2, 3,..., n dimensiones.
- Se puede elegir un Espacio 1D, 2D, o nD dividido en un número de subespacios (Celdas), con condiciones de frontera abierta o periódica.
- El conjunto de estados $\Sigma$ no necesita tener ninguna estructura algebraica adicional.
- La vecindad puede ser simétrica o no y puede incluir o no a la propia celda.
- La regla de evolución es una tabla o unas reglas.

Existen una gran cantidad de autómatas con diferentes reglas de evolución que fueron desarrollados para describir gran cantidad de fenómenos. Durante este trabajo tenemos como objetivo realizar una simulación de la dinámica de la lámina de plasma, mediante el uso de un autómata celular.

**Materiales y Métodos**

El modelo de la pila de arena, ha sido uno de los autómatas celulares más populares, porque fue el primero donde se apreció la conducta de criticalidad autoorganizada [*Bak et al. 1987, 1988*]. Muchos sistemas físicos reales muestran esta conducta, entre ellos: la lámina de plasma de la magnetosfera. El modelo de la pila de arenas ha sido usado con este objetivo en diversos trabajos [*Uriskii y Pudovkin, 1998; Uriskii y Semenov, 2000; Takalo et al. 2000. Uriskii et al. 2001*].

*Koselov y Koselova (2002 a, b),* aplicaron el modelo de la pila de arenas a la lámina de corriente para estudiar fundamentalmente las subtormentas magnéticas. En el último de sus modelos la lámina de plasma de la magnetocola es presentada como un arreglo bidimensional de celdas de 50×100, donde una frontera del arreglo está cerrada (la región nocturna terrestre donde se encuentra la lámina de corriente), en tanto las otras fronteras son abiertas. El estado de la celda con coordenadas (i; j) en el momento t está caracterizado por la energía almacenada $E_t(i ; j)$. Además existe un valor $C_t(i ; j)$ conectado a cada celda como una analogía con la conductividad de la ionosfera en el respectivo tubo de campo que interconecta diferentes regiones de la lámina de plasma con sitios en la ionosfera.

La entrada de energía al sistema corresponde al valor de la componente z del Campo Magnético Interplanetario (Bz). La energía penetra al sistema a través de la frontera del arreglo que corresponden a la primera y última fila de la matriz. Este valor es cambiado en dicho modelo a una velocidad de una celda por minutos por toda la frontera y a su vez la perturbación entra al sistema con esta misma velocidad, pero cambia proporcionalmente con la distancia a la frontera. Cuando el umbral se excede, las celdas pasan a un estado activo y cierta fracción de la energía almacenada. $\Delta E = E_t(i,j) - E_{min}$ se redistribuye entre cuatro de los vecinos más próximos de la celda:

$$E_{t+1}(i, j) = E_{min},$$
$$E_{t+1}(i+1, j) = E_t(i+1, j) + 0.25\Delta E,$$
$$E_{t+1}(i-1, j) = E_t(i-1, j) + 0.25\Delta E,$$
$$E_{t+1}(i, j+1) = E_t(i, j+1) + 0.4\Delta E, \qquad (1)$$
$$E_{t+1}(i, j-1) = E_t(i, j-1) + 0.1\Delta E,$$
$$donde: \Delta E = E(i, j) - E_{min}(i, j)$$
$$y\ E_{min}(i, j) = \begin{cases} kE_{max} \ para\ C_i(i, j) < C_{max} \\ 0\ para\ C_i(i, j) \geq C_{max} \end{cases}$$

Es costumbre llamar a tal reacción, avalancha, por la analogía con los procesos transcientes en el modelo de formación de una vertiente en una pila de arena [*Bak et al. 1988; Jensen, 1998*]. En contraste con algunos modelos [*Chapman., 1998*], es lógico asumir que la redistribución local de energía en la lámina de plasma lleva a una variación en la conductividad de la ionosfera en el respectivo tubo de campo magnético (por ejemplo, las irregularidades en la difusión de partículas debido al ángulo de pitch, es decir la precipitación de partículas por el cono de pérdida e ionización de gases atmosféricos). Esto se tiene en cuenta en el modelo de la siguiente manera:

$$C_{t+1}(i, j) = aC_t(i, j) + b \qquad (2)$$

Aquí, a < 1 tiene el sentido de un coeficiente de recombinación y el segundo término es determinado como:

$$b = \begin{cases} 0, para \to E_t(i,j) < E_{max} \\ E_{t+1}(i,j) - E_t(i,j), Para \to E_t(i,j) \geq E_{max} \end{cases} \quad (3)$$

El crecimiento de la conductividad de la Ionosfera sobre un cierto valor $C_{max}$, en un paso de tiempo, lleva a un cambio en las condiciones de redistribución de energía en la lámina de corriente (debido a la formación de las corrientes de campo alineadas). Ellos también consideraron este efecto en el modelo como una dependencia de $E_{min}$ con $C_t(i, j)$. Este valor de $E_{min}$ determina el fragmento de energía que puede redistribuirse en una celda activa en el próximo paso de tiempo. De forma similar [*Kozelov y Kozelova, 2002a*], asumen que el valor de $E_{min}$ es constante e idéntico en todas las celdas, sin embargo en un trabajo posterior estos autores [*Kozelov y Kozelova, 2002b*] lo toman variable.

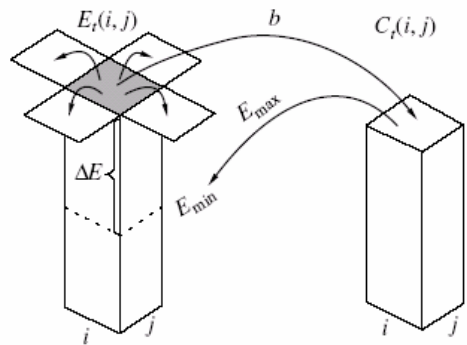

**Fig.1**- Representación esquemática de la realimentación positiva que cambia la conductividad en la ionosfera en una celda de la matriz que modela la lámina de plasma tomado de *Koselov y Koselova (2002b)*.

Los valores, $k < 1$ y $C_{max}$ son parámetros, de esta forma, la regeneración positiva surge entre la Magnetosfera y la Ionosfera características de cada celda. Esta regeneración positiva es mostrada esquemáticamente en la Fig. 1. El modelo numérico que ellos presentan tiene seis parámetros, dos de cuales —$E_{max}$ y $E_t(i,j)$— se consideraron de control. Los restantes parámetros modelados N, a, k, y $C_{max}$, se asume que son constantes. Este modelo es retomado por nosotros con algunos cambios para estudiar la interacción con la magnetosfera terrestre de nubes magnéticas, plasmoides y viento solar no perturbado. Aquí como parámetro de control del modelo, usaremos la componente Bz del Campo Magnético Interplanetario para ventanas temporales de viento solar del orden de un día y con una resolución temporal de 1 min.

Transformaciones al modelo.

Se asume la presencia de un arreglo matricial de 40×80 en analogía a la lámina de plasma, al igual que en los trabajos de *Takalo y Timonen, (1999); Uritskii y Pudovkin, (1998a), Consolini y De Michelis (2001), y Koselov y Koselova (2002a y b)*, donde cada celda de la matriz corresponde a un sitio en la lámina de plasma donde una línea de campo magnético puede interconectar este sitio con otro en la ionosfera (ver Fig.2).

Los portadores de carga se unen a las líneas de campo magnético y se precipitan en una zona de la ionosfera, aumentando la conductividad en la misma, lo cual tendremos en cuenta en el modelo, y estudiaremos dos tipos de transcientes o deposición de energía desde cada sitio en la lámina de corriente hasta la ionosfera. Estos transcientes se relacionarán con la ocurrencia de subtormentas magnéticas [*Koselov y Koselova 2002a y b*]. Los transcientes tipo 1 y tipo 2 se definen de la ecuación 1 en dependencia del valor que toma $E_{min}$. En el tipo 1 hay mayor energía involucrada en el proceso que en el tipo 2, debido al aumento de la conductividad en la ionosfera por encima de un nivel umbral. Se asume una mayor cantidad de columnas (80) que de filas (40) debido a que la lámina de corriente tiene forma rectangular alargada en la parte nocturna de la magnetosfera terrestre. Nosotros tomamos una menor cantidad de filas y columnas que en *Koselov y Koselova (2002b)* para ahorrar tiempo de cálculo, lo que no afecta la dinámica del modelo en su conjunto.

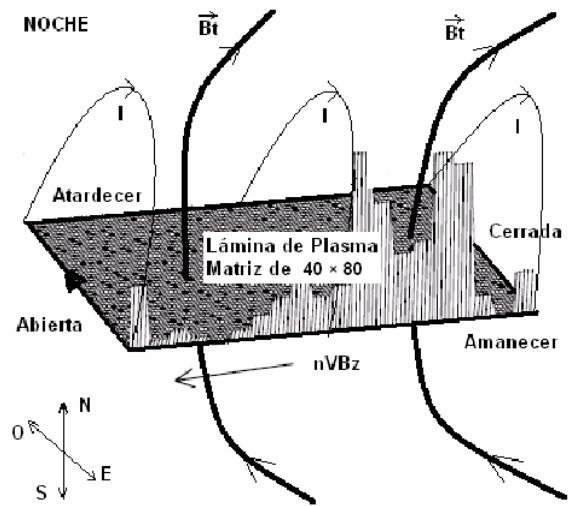

**Fig.2**- Esquema de la lámina de plasma modelada.

La energía que llega a la lámina de plasma está dada por valores reales de Bz tomados del viento solar por el satélite WIND. Se analizan casos reales de eventos importantes como son las nubes magnéticas, los plasmoides y el viento solar tranquilo [*Ojeda et., al 2005*]. Se toman ventanas de tiempo de aproximadamente un día de duración, con una discretización de 1 minuto. Los valores de entrada están dados por el módulo del valor de Bz, de la siguiente manera:

$$E_{input} = \begin{cases} |B_z| & si \ B_z < 0 \\ 0.1 B_z & si \ B_z \geq 0 \end{cases} \qquad (4)$$

Al asumir que para Bz > 0 una fracción de energía entra al sistema, tenemos en cuenta la energía que penetra por otros mecanismos que no es sólo la reconexión magnética con un Bz negativo. Es conocida la entrada de energía a través de la magnetosfera, debido a la interacción de tipo viscosa [*Axford y Hines, 1961*] y por la reconexión con Bz>0 [*Song et al., 2000*]. Asumir esta fracción de entrada de energía al sistema hace al modelo físicamente más real. Se supone la influencia de una velocidad finita de propagación en la matriz a lo largo de toda su frontera.

El valor de las celdas fronteras del arreglo $E_{input}$ (1; j) se cambiaron a lo largo de la frontera con la velocidad de 1cell / min , agregándose a cada celda de la frontera un valor de energía dE(1, j) = $E_{input}$/m en cada paso de tiempo, siendo m el número de columnas de la matriz (en *Koselov y Koselova (2002b)* se agrega toda la energía de una vez a todos los elementos de ambas fronteras de la matriz). La velocidad de propagación de la perturbación dentro de la matriz también se asumió que era de 1cell/min, y el valor de la perturbación disminuyó proporcionalmente a la distancia de la frontera (similar a *Koselov y Koselova 2002b*). La energía que llega a la otra frontera, se devuelve a la frontera de entrada debido a la corriente que circunda la magnetocola (Tail Current).

La reasignación local de energía en la magnetosfera causa un cambio local de conductividad de la ionosfera en el mismo tubo magnético (se precipitan las partículas, difundidas por el ángulo de pitch, en el Cono de pérdida (loss-cone) a lo largo del campo magnético, e ioniza los gases atmosféricos.). Tengamos en cuenta este efecto en el modelo, como fue presentado por *Koselov y Koselova (2002b)* (ver ecuación 2). Aquí a = 0.2 y se denomina « coeficiente de recombinación», por consiguiente la conductividad de la parte de la ionosfera de una celda depende de la historia de la celda. El segundo término se determina como en la ecuación 3. A su vez, si la conductividad de la ionosfera excede algún nivel $C_{max}$ entonces las condiciones de reasignación de energía en la lámina de plasma se cambian (a expensas de la formación de las líneas de corrientes del campo). En el modelo nosotros tendremos en cuenta esta influencia como la relación de $E_{min}(i,j)$ con $C_t(i,j)$. Este valor $E_{min}(i,j)$ determina, qué parte de energía puede reasignarse en una celda activa al paso de tiempo siguiente (ver ecuación 1). Ahora, k<1 (k = 0.75) y $C_{max}$ es igual al valor medio que existe en toda la matriz conductividad, hecho que hace diferente al que toman [*Koselov y Koselova 2002b*], donde $C_{max}$=5.

Existe un valor máximo de energía ($E_{max}$) sobre el cual una celda se torna activa [*Bak, 1997*]. Este valor está dado por la energía que penetra desde el viento solar y por tanto es variable en la evolución del sistema. Después de la redistribución de energía desde el viento solar, pueden existir varias celdas activas en el sistema y ocurrir transiciones internas en el mismo, siendo éstas mucho más veloces que las externas, por tanto en el orden de un minuto el sistema tendrá tiempo de relajarse completamente antes de la llegada de la próxima perturbación externa desde el viento solar.

Puede haber en el sistema más de una celda activa, y a partir de cada una de ellas surgirá una avalancha de diferente tamaño y duración. Las fronteras siempre son un problema a la hora de programar el modelo. En este caso teníamos tres fronteras abiertas y una cerrada, algo atípico en un autómata celular. La frontera cerrada era la columna 1 que corresponde al lado de la Tierra de la lámina de corriente. Por tanto, si una de estas celdas se activa, sólo se entregará energía a tres vecinos y no a cuatro, para el caso de las celdas de la última columna que corresponde al lado más lejano de la lámina de plasma a la Tierra, se entregará energía a tres vecinos y la otra se expulsará del sistema, en analogía a la energía que verdaderamente es devuelta al viento solar en el extremo lejano de la magnetocola. Las otras dos fronteras, de la fila 1 y f, son periódicas, debido a la corriente que va desde el lado diurno al nocturno de la lámina de corriente (Neutral Sheet Current).

## Resultados y discusión

Uno de los objetivos de nuestro trabajo es estudiar más detalladamente la influencia de la variable Bz en las perturbaciones geomagnéticas a nivel global, partiendo del modelo de *Koselov y Koselova (2002b)*, pero con algunas transformaciones. Estos autores hacen una simulación de la realidad física del comportamiento de la lámina de plasma, coincidiendo con los resultados experimentales mostrados en los Keogramas y en el trabajo a partir de datos experimentales de *Lui et al. (2000)*.

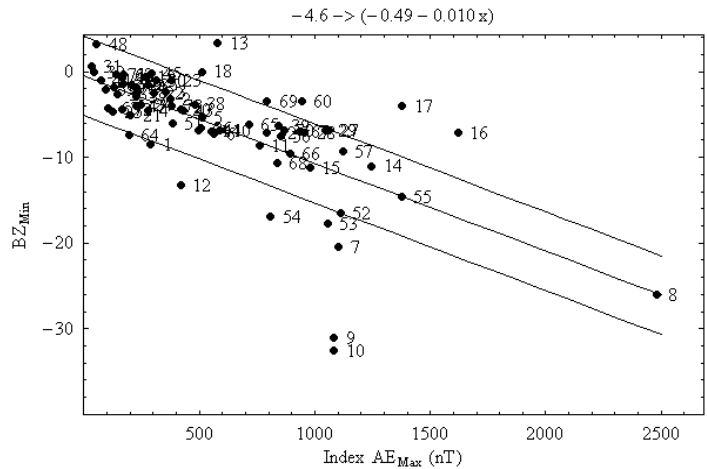

**Fig.3**- Se muestra la relación entre el valor de Bz mínimo y el máximo valor del índice AE para ventanas temporales de 3 horas tomando de referencia la serie temporal del Bz que se utiliza como parámetro de control al modelo del autómata celular estudiado. La ecuación de la recta es Bz= -0.01AE -0.49.

Ya en el apéndice anterior se explicó de forma general las transformaciones que realizamos al modelo de *Koselov y Koselova (2002b)*, ahora a modo de discusión, queremos profundizar en este aspecto:

- La entrada de energía al sistema no se realiza por ambas fronteras del sistema desplazándose hasta el interior. Ahora la energía penetra a la matriz por el lado del amanecer y se mueve hasta las celdas interiores siguiendo el sentido de la corriente neutral de la lámina (Neutral Sheet current). Cuando llega a la frontera del atardecer se incorpora en el próximo paso de tiempo a la frontera del amanecer debido a la corriente de la cola (Tail Currents). En nuestro caso, se deposita el total de esta energía que llega cada minuto (ver ecuación 4) en fracciones iguales del total (dE = $E_{input}/80$), desde la primera columna, hasta la última de la primera fila y a una velocidad de una celda por minuto, considerando que la energía no llega al mismo tiempo al principio de la lámina de plasma (que se encuentra aproximadamente a 8 $R_T$(radios terrestres)) que al final (a mas de 100 $R_T$). *Koselov y Koselova (2002b)* en el mismo minuto que llega $E_{input}$ es entregada en fracciones iguales a toda la frontera.

- El valor de $E_{max}$ para que se produzca una avalancha debe ser mayor o igual que el valor medio de toda la ventana de tiempo de $E_{input}$ que conocemos de antemano.

$$E_{\max} = \begin{cases} \overline{E}_{input} & si \quad \mathrm{E}^{\min}_{input} \leq \overline{E}_{input} \\ \mathrm{E}^{\min}_{input} & si \quad \mathrm{E}^{\min}_{input} > \overline{E}_{input} \end{cases} \tag{5}$$

$\mathrm{E}^{\min}_{input}$, es la energía que comenzó a entrar en ese minuto por la primera fila.

*Koselov y Koselova (2002b)*, toman el valor de $E_{\max} = \mathrm{E}^{\min}_{input}$, por lo que en el sistema ocurren grandes avalanchas para un Bz próximo a cero y casi ninguna para un Bz muy negativo, lo que difiere del comportamiento real del sistema y de los modelos MHD que describen al mismo.

El resultado obtenido por nosotros en la figura 3, en el cual se grafica el valor mínimo real del Bz para una ventana temporal de 3 horas contra el máximo valor del índice AE en esa ventana para los 10 casos estudiados, también se aparta del propuesto por *Koselov y Koselova (2002b)*. Si partimos de que un valor grande del índice AE implica mayor actividad auroral debido a la deposición de energía desde la lámina de plasma hasta la ionosfera auroral, ocurriendo un aumento en la circulación de las corrientes aurorales (Pedersen Curren y Hall Current), el modelo debiera responder aumentando el número y tamaño de las avalanchas, por lo cual, la forma que emplean *Koselov y Koselova (2002b)* no responde a la física del sistema.

La figura 3 muestra lo geoefectivo que es el Bz < 0 del Campo Magnético Interplanetario para la incorporación de partículas a la magnetosfera terrestre y que generan subtormentas y tormentas magnéticas. La reconexión magnética una vez más se muestra como un mecanismo muy efectivo en el intercambio de partícula entre dos plasmas guiados por líneas de campo magnético. En la literatura se habla de la importancia del Bz negativo para que ocurran subtormentas [*Zhou y Tsurutani, 1999*] pero no hemos encontrado ninguna mención en la literatura de que a menor Bz, más geoefectivo sea. Por tanto, lo anterior parece ser un nuevo resultado encontrado.

- El valor de la máxima conductividad ($C_{\max}$) para disparar los transcientes de primer tipo en el modelo de *Koselov y Koselova (2002b)* fue tomado como constante e igual a 5 (ver ecuación 2.5) sin explicar los motivos. Cuando se realiza una simulación esto es válido, de hecho en la mayor parte de los autómatas celulares que se usan en la literatura se procede de esta forma [*Bak, Tang y Wiesenfeld, 1988*]. Nosotros decidimos calcular este valor para cada paso de tiempo como el valor promedio de la conductividad en toda la matriz y si bien tampoco somos capaces de poner un valor que esté referenciado en otro trabajo científico anterior, sí pensamos esté mas cerca de la realidad física del problema, puesto que el mismo adquiere información constante del sistema.

- La duración de las perturbaciones internas cuando se dispara una avalancha es otra gran diferencia. Ellos no toman el criterio que se usa en el modelo de pilas de arenas "cuando cae un grano no se vuelve a depositar otro hasta que dejen de existir avalanchas en el sistema". Ésta forma de evolución del sistema hace posible que se observen leyes de potencias y que en el modelo de pilas de arena se observe la criticalidad autoorganizada (SOC). En cada deposición de energía en la evolución de una avalancha pasan 10 segundos (para *Koselov y Koselova 2002b*), de manera que cuando la avalancha tenga tamaño 6, a cada celda de la matriz se le agrega una cantidad de energía producto de la perturbación externa que entra por sus fronteras. Ellos no justifican el por qué toman 10 seg en tales transiciones internas, sólo plantean que éstas son más rápidas que las externas y tampoco explican con claridad si a las celdas que están involucradas en la avalancha se le agrega la energía perteneciente a la perturbación externa. Nosotros no contamos este tiempo y mientras exista una celda activa en el sistema no agregamos el valor de la energía externa a las celdas del sistema. Decidimos hacerlo así debido a que existe mucha evidencia experimental [*Uritskii y. Pudovkin, (1998b, 2000 y 2001), Consolini y De michelis 2001*] de que las subtormentas magnéticas responden a descargas de energía de un sistema con SOC.

La simulación realizada por *Koselov y Koselova (2002b)* y que hemos transformado es completamente determinista y responde a un impulso externo caótico [*Takalo y Timonen 1999*] con valores reales (Bz), mientras que los modelos clásicos de autómatas celulares se generan a partir de la introducción de números aleatorios para la evolución del sistema [*e. g. Bak, Tang y Wiesenfeld, 1988*], lo que en nuestra opinión es una gran ventaja. Por otra parte, coincidimos plenamente con *Koselov y Koselova (2002b)* cuando dicen: "La reasignación local de energía en la magnetosfera causa un cambio local de conductividad de la ionosfera en el mismo tubo magnético (se precipitan las partículas, difundidas por el ángulo de pitch, en el cono de perdida (loss-cone) a lo largo del campo magnético, e ioniza los gases atmosféricos.)" y lo expresan en las ecuaciones 1, 2 y 3. Nosotros no realizamos ninguna transformación a este criterio.

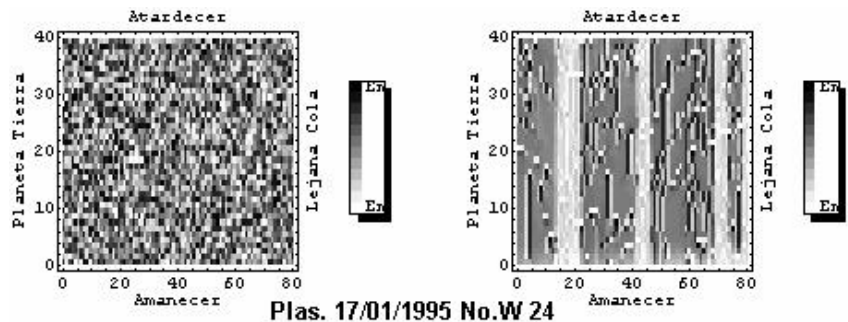

**Fig. 4-** En el panel de la izquierda se muestra la energía que inicialmente se le dio a la matriz en cada celda (se usó un generador de números aleatorios). A la derecha se aprecia cómo queda cada celda después de la corrida. Se pueden ver patrones ordenados, mostrando que en el sistema modelado existe auto-organización.

Análisis de los resultados

Se usaron como parámetro de control al modelo 10 juegos de datos de la componente z del Campo Magnético Interplanetario medidos por el satélite WIND. Los eventos corresponden a 2 plasmoides, 4 ventanas de viento solar no perturbado tomado un día antes de cada nube magnética y dichas nubes (4) respectivamente. En la figura 4 se muestra la distribución de energía en cada sitio de la matriz desde una vista superior (se obtuvo un resultado similar para cada uno de los 10 eventos estudiados). El color varía desde el negro al blanco representando los valores de energía que existen en cada sitio desde el mínimo al máximo. En el panel de la izquierda se muestra la matriz de partida que se obtuvo a partir de un generador de números aleatorios (Gaussiano). En el panel de la derecha mostramos la distribución de energía de la matriz después de una corrida con un juego de datos de Bz. El sistema en todos los casos tiene patrones organizados de colores en la distribución de energía. Lo anterior evidencia que el sistema logra autoorganizarse a sí mismo bajo la influencia de un impulso externo caótico [*Takalo y Timonen 1999*].

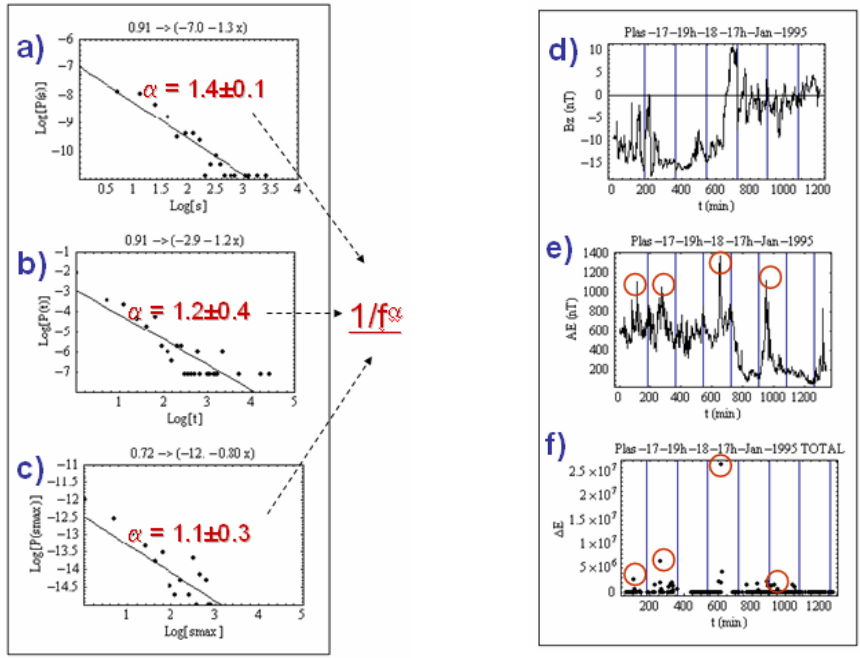

**Fig. 5-** Resultados para una corrida con entrada de datos del Bz correspondiente a un plasmoide (17.01.1995). En los gráficos de la izquierda se ajusta una recta en busca de leyes de potencias: a) Gráficas de probabilidades de ocurrencia de una avalancha de determinado tamaño, b) Probabilidad del tiempo de espera (Waiting Time) desde que se inicia una avalancha de determinado tamaño e inicia otra, c) Probabilidad de la cantidad máxima de celdas activas en cada avalancha y qué común esto es a lo largo de la evolución del sistema. En los gráficos de la derecha mostramos: d) ventana de tiempo del parámetro de control del modelo reportado cada minuto (Bz) por el satélite WIND, e) Índice AE, f) Energía total disipada en los transcientes de tipo 1 y 2.En todos los gráficos las ventanas temporales enmarcan un intervalo de tres horas. Cada valor de energía se disipa a lo largo de una avalancha en f). Los picos en el valor de energía en muchos casos coinciden con los picos en los valores del índice AE como se ve en e) y f).

De lo anterior no podemos afirmar que estemos en presencia de un sistema con criticalidad autoorganizada, para ello, como se ha dicho con anterioridad a lo largo de este trabajo, debemos observar la presencia de funciones de distribución que muestren leyes de potencias en los tamaños y tiempos de espera de las avalanchas, con exponentes $1<\alpha<2$. Si encontramos leyes de potencias podemos ir mas allá de decir que existe invarianza de escala, pues en el modelo se toma el mismo criterio que en el modelo de pilas de arenas, donde el tiempo de agregar un grano es mucho mas grande que el tiempo de ocurrencia de las avalanchas. Nosotros agregamos energía al sistema cada un minuto, pero nos aseguramos que ya no existirían celdas activas en la matriz.

En la figura 5, panel c) representamos la probabilidad de la cantidad máxima de celdas activas en cada avalancha y qué común es a lo largo de la evolución del sistema. Se graficó en ambos ejes el logaritmo. La pendiente de la mejor recta es igual al exponente $\alpha$ de la ley de potencia ($1/f^{\alpha}$). Se puede reportar en general un exponente $\alpha = 1.1 \pm 0.2$ entre los 10 casos estudiados. En el panel a) se muestran el gráfico de probabilidades de ocurrencia de una avalancha de determinado tamaño. Se puede reportar un exponente $\alpha = 1.4 \pm 0.1$ entre los 10 casos estudiados. En los paneles b) se observan las probabilidades del tiempo de espera desde que se inicia una avalancha de determinado tamaño e inicia otra. Se puede reportar un exponente $\alpha = 1.8 \pm 0.4$ a partir de los 10 casos estudiados.

De los dos párrafos anteriores podemos sacar una conclusión importante: El sistema al final de una corrida muestra patrones organizados en la distribución de energía en cada celda, lo que unido a las leyes de potencias que se observan en el número máximo de celdas activas, tamaño y tiempo de espera en las avalanchas dan muestra de un sistema con criticalidad-autoorganizada (SOC), independientemente del tipo de evento. Esto está en correspondencia con algunos trabajos que se han hecho en este sentido, que han estudiado el comportamiento SOC de la magnetosfera [*Chang 1999a y b, Chapman et. al., 1998, Drossel y Schwabl 1992*].

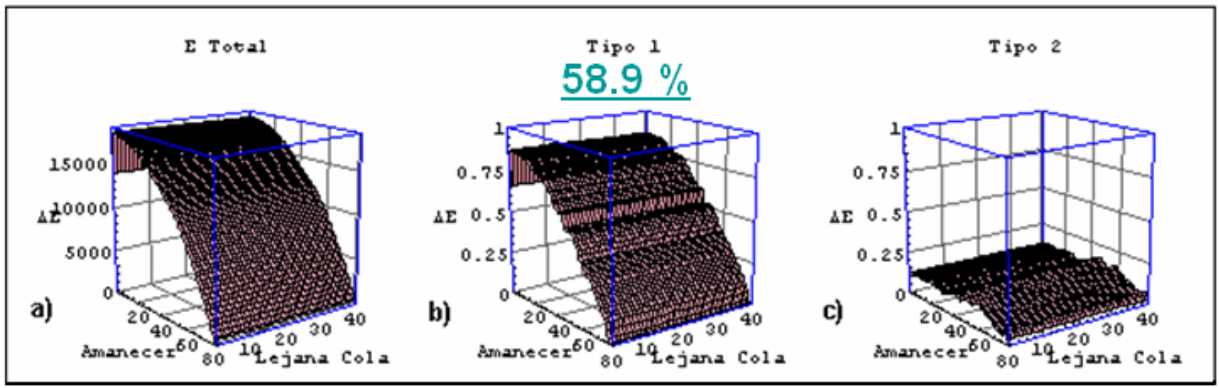

**Fig. 6-** Representación del arreglo matricial que muestra la energía disipada en cada celda durante una corrida del modelo: a) La energía total disipada, b) Sólo para transcientes del primer tipo (58.9%), c) Transcientes de segundo tipo.

Producto de la reasignación local de energía podemos separar dos tipos de deposiciones de energía o de transcientes desde la lámina de plasma hasta la ionosfera auroral. Llamamos transcientes de primer tipo (o tipo 1) aquellos que se disparan cuando la conductividad en su correspondiente sitio en la ionosfera (C(i,j)) es mayor que un valor umbral (Cmax). Los transcientes de segundo tipo (o tipo 2) son aquellos donde ocurre lo contrario. A partir de las ecuaciones 1, 2 y 3 se puede deducir que los tipo 1 son más intensos que los tipo 2. Ambos transcientes pueden ser estudiados a partir de los resultados mostrados en la figura 6.

Los tubos de plasmas son los encargados de retornar el plasma hasta el interior de la magnetosfera desde la lámina de plasma según observaron experimentalmente *Sergeev V.A. et. al (2000)*, ellos obtuvieron que estos flujos se establecen fundamentalmente desde la mitad de la lámina de plasma a 40 $R_T$ y demoran 10 minutos hasta 6.6 $R_T$. En los 10 casos estudiados los transcientes de tipo 1 se presentaron desde 54 al 67% del total, lo cual evidencia un menor número de transcientes de segundo tipo. Cuando se dispara una avalancha ocurren en la lámina de plasma el disparo de transcientes de primer tipo y de segundo tipo. Para avalanchas del mismo tamaño, cuando más tipos 1 ocurren, mayor energía se deposita hasta la ionosfera auroral. A mayor energía que se disipe en cada avalancha, mayor debe ser la subtormenta magnética asociada a ella.

En el panel a) observamos que la energía total disipada en la lámina de plasma es mayor en la parte cercana al planeta Tierra y decrece a medida que nos apartamos hasta la lejana cola. Una característica de las subtormentas magnéticas es que aumentan el flujo de partículas hasta el óvalo auroral provocando que éste se expanda [*Newell et al., 1998*]. Esta expansión significa que los tubos de plasmas que ahora llegan a la zona auroral bajan a menores latitudes y según nuestros resultados, tal parece que estos tubos de plasmas están comunicados entonces con una región cercana de la lámina de plasma. Luego, cuando se produce una gran eyección coronal de masa (CMEs) en el Sol, ésta al interactuar con el campo magnético de la Tierra comprime la magnetosfera por el lado diurno y comprime también los lóbulos de la cola, provocando que esta última se vuelva más larga.

Los transcientes tipo 1 son más comunes en la zona cercana de la magnetosfera (ver panel b), en tanto los transcientes tipo 2 tienen un máximo transversal en la mitad de la lámina de plasma (ver panel c). Quizás éstos fueron los transcientes observados experimentalmente a $40R_T$ por *Sergeev V.A. et. al (2000 )* haciendo un estudio multisatelital. Por último, observamos que en todos los paneles como el a) en los 10 casos estudiados, hay eventos donde se deposita mayor energía que en otros, como ocurre, por ejemplo, en la nube magnética que arribó el día 08 de febrero de 1995, donde el valor máximo de energía disipada es cercana a 15000 unidades de energía, perturbando el índice $Dst_{min}$ = -80 nT y el $AE_{max}$ = 1622 nT. Analizando una ventana temporal de la misma duración un día antes, cuando existía un viento solar no perturbado ($Dst_{min}$ = -32 nT y el $AE_{max}$ = 422 nT), se disipó un valor menor que el caso anterior, sólo que la cifra de 5000 unidades de energía. En los restantes pares de nubes magnéticas, con su correspondiente viento solar no perturbado un día antes, se observó esta correlación.

En el plasmoide correspondiente al día 17 de enero de 1995 (ver figura 6, $Dst_{min}$ = -95 nT y $AE_{max}$ = 1378 nT) y el plasmoide del día 6 de abril de 2000 ($Dst_{min}$ = -321 y $AE_{max}$ = 2483 nT) los valores de energía disipados fueron de 35 000 y 45 000 unidades de energía respectivamente, observándose que a mayor AE, mayor es la energía disipada. Podemos concluir que las avalanchas cubren los gastos energéticos necesarios para que se disparen las subtormentas geomagnéticas reales, a partir de analizar los datos del índice geomagnético AE.

**Conclusiones**

1- La llegada a la magnetosfera terrestre de un viento solar con un Bz negativo es la condición fundamental para la ocurrencia de subtormentas magnéticas. Mientras mayor sea el valor modular del Bz negativo mayor es el valor del índice AE y activa una fuerte subtormenta magnética.

2- El sistema al final de una corrida muestra patrones organizados para la distribución de energía en cada celda, y responde matemáticamente a leyes de potencias al estudiar la función de distribución de probabilidad según el número máximo de celdas activas, su tamaño y por tiempos de espera. El orden de los coeficientes obtenidos en cada caso analizado muestra evidencias de un sistema con Criticalidad Auto-Organizada (SOC), independientemente del tipo de evento.

3- La energía disipada en las avalanchas del modelo a lo largo de la evolución de cada evento estudiado y comparado con otros, cubren los gastos energéticos necesarios para que se disparen las subtormentas geomagnéticas reales a partir de analizar los datos del índice geomagnético AE.

4- El disparo de transcientes de energía a través de los tubos de plasma desde la lámina de plasma hasta la ionosfera auroral y que causan las subtormentas geomagnéticas, ocurren con mayor frecuencia en la primera mitad de la lámina de plasma, teniendo un máximo los transcientes de tipo II en la mitad de dicha lámina.

## Referencias